%% file: paper.tex
\documentclass[10pt,conference]{IEEEtran}

\usepackage{url}
\usepackage{graphicx}
\usepackage[]{color}
\usepackage{xspace}
\usepackage{flushend}
\usepackage{hyperref}
\usepackage[numbers]{natbib}
\usepackage{listings} 
\usepackage{rotating} 
\usepackage{caption} 
\usepackage{subcaption} 
\usepackage{booktabs} 
\usepackage{amssymb}
\usepackage[inline]{enumitem}
\usepackage{amsmath,amssymb,amsfonts}
\usepackage{textcomp}
\usepackage{xcolor}

\usepackage{amsthm}
\theoremstyle{definition}

\input macros

\begin{document}

\title{Search-Based Software Re-Modularization:\\A Case Study at Adyen}

\author{\IEEEauthorblockN{Casper Schröder, Adriaan van der Feltz}
\IEEEauthorblockA{\textit{Adyen N.V.} \\
Amsterdam, the Netherlands\\
\{casper.schroder, adriaan.vanderfeltz\}@adyen.com}
\and
\IEEEauthorblockN{Annibale Panichella, Maurício Aniche}
\IEEEauthorblockA{\textit{Delft University of Technology}\\
Delft, the Netherlands \\
\{A.Panichella, M.FinavaroAniche\}@tudelft.nl}
}

\maketitle

\input sec-abstract


\sloppy 

\input{sec-introduction}
\input{sec-related-work}

\input{sec-case-study}

\input{sec-approach}

\input{sec-research-part1}

\input{sec-research-part2}

\input{sec-challenges-ooportunities}

\input{sec-ttv}

\input{sec-conclusions}

\renewcommand*{\bibfont}{\footnotesize}

\bibliographystyle{IEEEtranN}
\bibliography{IEEEabrv,paper.bib}

\end{document}

%% file: macros.tex
\newcommand{\adyen}{\emph{Adyen}\xspace}

\newcounter{observations}
\newcommand{\observation}[1]{\refstepcounter{observations} \vspace{2mm} \textbf{Observation \theobservations: #1}}

\newcounter{challenges}
\newcommand{\challenge}[1]{\refstepcounter{challenges} \vspace{2mm} \textbf{Challenge \thechallenges: #1}}

\newtheorem{thm}{Definition}[subsection]

%% file: sec-abstract.tex
\begin{abstract}
Deciding what constitutes a single module, what classes belong to which module or the right set of modules for a specific software system has always been a challenging task. The problem is even harder in large-scale software systems composed of thousands of classes and hundreds of modules. Over the years, researchers have been proposing different techniques to support developers in re-modularizing their software systems. In particular, the search-based software re-modularization is an active research topic within the software engineering community for more than 20 years.

This paper describes our efforts in applying search-based software re-modularization approaches at \adyen, a large-scale payment company. \adyen's code base has 5.5M+ lines of code, split into around hundreds of modules. We leveraged the existing body of knowledge in the field to devise our own search algorithm and applied it to our code base. Our results show that search-based approaches scale to large code bases as ours. Our algorithm can find solutions that improve the code base according to the metrics we optimize for, and developers see value in the recommendations. Based on our experiences, we then list a set of challenges and opportunities for future researchers, aiming at making search-based software re-modularization more efficient for large-scale software companies.

\end{abstract}

%% file: sec-introduction.tex
\section{\label{cha:intro}Introduction}

The existing body of knowledge on how to structure a software system in such a way that its evolution and maintenance can be done in effective and sustainable ways is extensive; it ranges from the canonical works by Parnas on software modularization~\cite{parnas1972criteria}, to Booch et al.'s view on object-oriented software design~\cite{Booch2008}, to modern software design techniques, such as Domain-Driven Design (DDD)~\cite{evans2004domain}, to microservices~\cite{newman2015building}. 
Nevertheless, designing large-scale software systems is still an important and practical challenge for software development teams. 

In this paper, we highlight the challenges of modularizing a large-scale software system. Modularity is the idea of partitioning a (often large) software system into different components to reduce its overall complexity~\cite{myerscomposite}.
However, deciding what constitutes a single module, what classes belong to which module or the right set of modules for a specific software system is challenging. The challenge might be even harder at the initial phases of the software development process, as little is known about the application itself~\cite{Booch2008}.
Finally, the way contemporary large software companies (such as \adyen, the industry partner that serves as a case study in this research) work and divide the development responsibilities also augment the challenge: modules evolve in fast paces, are developed by different software teams, and any possible intersections between modules (e.g., does this responsibility belong to module A or B?) are solved ad-hoc. 

Modularization of code has been a software engineering research topic for more than 20 years. We see works focusing on, e.g., proposing code metrics that aim at detecting modules that are not cohesive or are too much coupled (e.g., ~\cite{Allen2001, Mancoridis1998, Bavota2012strucsem}), and automated detection of modularization anti-patterns (e.g.,~\cite{xiao2016identifying}). Closely related to this research, we also see a large effort from the community in modeling software (re-)modularization as an optimization problem and employing search-based techniques to find candidate solutions (e.g.,
~\cite{Kwong2010, Kessentini2011, Mitchell2006, Praditwong2011,  Mkaouer2016, Abdeen2013, Ouni2012, Ghannem2017, Mahouachi2018, Fadhel2012, Bavota2012interact, Alizadeh2018, Harman2002, daSilva2020}). 
The different studies have relied on different software quality metrics, e.g., coupling, cohesion, number of modules and clusters, effort, package size, number of changes, and semantic cohesiveness. They also use different optimization algorithms, e.g., hill climbing, vanilla GA, NSGA-II, interactive GA, NSGA-III, and MOEA. However, while current results are promising and the accuracy of these techniques has been evolving over time, we still only have little evidence of how these techniques can work in large-scale enterprise software systems.

In this paper, we describe an industrial case study on the effectiveness of search-based approaches for software re-modularization at \adyen, a large-scale enterprise software system composed of millions of lines of code and hundreds of developers. 
We also proposed a new quality metric (search objective) that takes transitive coupling and building time as part of the search problem, an objective that is perceived as fundamental in practice by the \adyen developers. We present our findings in two phases. First, we analyzed the improvements we obtain by looking at the solutions in the Pareto front. Second, we presented a set of the results to different developers, experts in the modules that the improvement was suggested and collect their perception.

Our results show that 
\begin{enumerate*}[label=(\roman*)]
    \item existing search-based approaches scale to code bases with similar sizes (5.5M+ lines of code), 
    \item solutions present significant improvements as measured by the optimization metrics used as search objectives,
    \item the novel metric that captures transitive dependencies and their impact on module caching is a useful metric for the re-modularization problem, and
    \item developers see value in the recommended refactoring operations. However, while the generated solutions identify classes in wrong modules often accurately, the recommendation of which module to move it to is often imprecise.
\end{enumerate*}

This paper makes the following contributions:

\begin{enumerate}

\item A case study on the effectiveness of multi-objective evolutionary optimization approaches to the software (re-)modularization problem on a large-scale enterprise software system.

\item A new optimization quality metric  (search objective) that takes into consideration transitive dependencies and build cost that can be used as part of the search process for better (re-)modularization solutions.

\item A set of challenges and opportunities for future work on search-based software re-modularization, derived from our experiences at \adyen.

\end{enumerate}

%% file: sec-related-work.tex
\section{Related Work}
\label{cha:related}

In Table~\ref{tab:related-work}, we list the different papers in the search-based re-modularization line of research that inspired us, grouped by the algorithms they apply and the metrics they optimize for.

Overall, we observe a vast mixture of different algorithms being applied, with vanilla genetic algorithms and the NSGA family being among the most popular ones. We also observe that coupling and cohesion metrics are often measured in different ways and are the most commonly used objectives.

Meta-studies in software refactoring seem to observe similar trends. \citet{Ebad2011} compare different approaches to the modularization problem. The authors created a framework that measures approaches based on the given packaging goal, underlying principle, input artifact, internal quality attribute, search algorithm, fitness function, scalability, soundness, practicality, and supportability. 
They observe the following aspects:
\begin{enumerate*}
\item most research sees packaging as an optimization problem,
\item Most research maximizes intra-package cohesion and minimize inter-package coupling,
\item most research uses genetic algorithms as their search methods,
\item all approaches are graph-based,
\item most use source code as the input artifact,
\item the selection of parameters of the heuristic search method is very crucial (due to local optima),
\item most approaches are scalable,
\item papers do not often provide tools that are easy to use (according to authors, only a single paper fit this criteria)
\end{enumerate*}


In this paper, we take advantage of the already extensive body of knowledge in the field, and apply it to a large-scale enterprise system. More specifically, we:
\begin{enumerate*}[label=(\roman*)]
\item evaluate whether the approaches proposed by the paper are indeed scalable, by applying it to a 5.5M+ code base (we note that most research deals with projects in the scale of thousands of lines of code, not millions of lines of code),
\item measure the perception of developers regarding the recommendations (we note that most research currently evaluates their results by means of either controlled or quantitative experiments, but not with the actual developers of the software systems).
\end{enumerate*}


\input tab-related-work

%% file: tab-related-work.tex
\begin{table}
\caption{Related work in search-based software re-modularization, grouped according to their algorithm and optimization variables of choice.}
\label{tab:related-work}
\begin{tabular}{p{4cm}p{4cm}}
\toprule
\textbf{Search algorithms} & \\
\midrule 
K-cut & \citet{Jermaine1999} \\
Clustering & \citet{KookLee2002} \\
Hill climbing & \citet{Mitchell2003, Mitchell2006, Mitchell2007}, \citet{Marx2010} \\
Genetic Algorithm & \citet{Harman2002}, \citet{Mitchell2003, Mitchell2006, Mitchell2007}, \citet{Kwong2010}, \citet{Praditwong2011} \\
Min-cut algorithms & \citet{Marx2010} \\
NSGA-II & \citet{Ouni2012}, \citet{Abdeen2013}, \citet{Mahouachi2018} \\
Interactive NSGA-II & \citet{Bavota2012interact} \\
NSGA-III & \citet{Mkaouer2015} \\
Exhaustive search & \citet{Marx2010} \\
 \midrule
\textbf{Search objectives} & \\
\midrule
Cohesion (intra-edges) & \citet{Jermaine1999}, \citet{Harman2002}, \citet{Mitchell2003, Mitchell2006, Mitchell2007}, \citet{Kwong2010}, \citet{Praditwong2011}, \citet{Mahouachi2018} \\
Cohesion (functionality similarity) & \citet{KookLee2002} \\
Cohesion (Common Closure Principle, Common Reuse Principle) & \citet{Abdeen2013} \\
Coupling (inter-edges) & \citet{Jermaine1999}, \citet{Harman2002}, \citet{Mitchell2003, Mitchell2006, Mitchell2007}, \citet{Kwong2010}, \citet{Praditwong2011}, \citet{Mahouachi2018} \\
Coupling (any type of interaction) & \citet{KookLee2002} \\
Coupling (Acyclic Dependencies Principle) & \citet{Abdeen2013} \\
Number of modules & \citet{Harman2002} \\
Functional performance & \citet{Kwong2010} \\
Number of broken dependencies & \citet{Marx2010} \\
Characteristics of the clusters & \citet{Praditwong2011} \\
Semantic similarity & \citet{Ouni2012}, \citet{Mkaouer2015}, \citet{Mahouachi2018} \\
Method calls and field accesses & \citet{Ouni2012} \\
Effort required & \citet{Bavota2012interact}, \citet{Mahouachi2018} \\
Package size & \citet{Abdeen2013} \\
Number of required changes & \citet{Abdeen2013}, \citet{Mkaouer2015}  \\
\bottomrule
\end{tabular}
\end{table}

%% file: sec-case-study.tex
\section{The Case Study}
\label{cha:case-study}

In this section, we give an overview of the business, size, and scale of our industry partner, and we list the challenges and assumptions that emerge when applying search-based software re-modularization techniques to the case study.

\subsection{Our industry partner: \adyen}

\adyen is a payment service provider, gateway, acquirer, and point of sale terminal service\footnote{\url{https://www.adyen.com/}}. The company was founded in 2006 and has grown rapidly, totaling over 1,400 employees as of August 2020. Most of the code is part of a monolithic repository, consisting of 5.5M+ lines of code. Due to the emphasis on uptime, performance, and growth potential for this type of business, good code quality, and code structure quality are essential. A code base of this size and speed of growth brings additional challenges when compared to smaller code bases.

\subsection{Definition of modules}

\adyen's code base is divided into around hundreds of modules, each module having a different business responsibility. In this paper, a module can be interpreted as a package comprised of one or more Java classes. In this paper, we aim at identifying classes that currently reside in the wrong module, and recommend which module to move the class to.

\subsection{Challenges of the scale}

\emph{Size and complexity of the code base.} The most obvious challenge coined by this case study is the size and complexity of the code base. As we mentioned, \adyen's code base is composed of 5.5M+ lines of code, split into hundreds of modules. This implies a natural increase in the input that the chosen algorithm will take and its time to find the optimal output. 

\vspace{2mm}
\emph{The impact of transitive dependencies.} The large number of different modules emphasizes another aspect: the impact of transitive coupling.
The transitive coupling can be described as follows: if a given module M1 depends on a given module M2, and module M2 depends on given module M3, M1 is, therefore, transitively dependent on M3. Given the large number of modules and their existing dependencies
, any refactoring that is suggested by the approach may profoundly impact the transitive dependencies. Moreover, moving a class from one module to the other may even affect build and compilation time. For example, suppose that the tool recommends some class C to move from module A to module B. All modules that depend on class C will now have to depend on module B. For some modules, this will mean an extra dependency. If C is a class that changes very often, module B (and all its recursive dependencies) will need to be re-compiled often. Companies that develop large-scale software systems like \adyen often rely on module caching to avoid the recompilation of the entire code base; any refactoring suggestion should therefore take the compilation time, and cache breaks into consideration. As we explain before, we use transitive dependencies and cache breaking as an optimization variable.

\vspace{2mm}
\emph{A full refactoring is not possible in practice.} Approaches that suggest completely new ways of modularizing the entire software system are bound not to be used at \adyen. Given the already large size of the code base, a full (or large) refactor is considered highly risky. Thus, any proposed approach should be able to propose refactoring solutions in small doses, i.e., they should be as small as possible.

%% file: sec-approach.tex
\section{The approach}
\label{cha:approach}

In this section, we discuss the design decisions we made in the search algorithm we propose to generate re-modularization refactoring suggestions at \adyen. 
In a nutshell, we formalize the re-modularization problem as follows:

\begin{thm} 
\label{def:problem}
\textit{Let $C=\{C_1, \dots, C_n\}$ be the set of classes in a software system, and let  $M=\{M_1 \dots, M_k\}$ be the set of existing software modules. Find a set of move class operations $X=\left\{ C_i \rightarrow  M_j, \dots, C_h \rightarrow  M_k\right\}$, $C_i \rightarrow  M_j$ denoting that the class $C_i$ is moved to the module $M_j$, that optimizes the following four search objectives:}
\begin{equation}
\centering
\left\{
\begin{array}{l}
  min f_1(X) = \textrm{coupling}(C, M, X) \\
  max f_2(X) = \textrm{cohesion}(C, M, X) \\
  min f_3(X) = |X| \\
  min f_4(X) = \textrm{cost-of-transitive-dependencies}(C, M, X)
\end{array}
\right.
\end{equation}
\textit{with the constraint that there is no circular dependencies across the modules $M$ after the refactoring (as otherwise it is not possible to build the software).}
\end{thm}

In the following sub-sections, we describe the problem representation, the search objectives, and the optimization algorithm we used in our approach in detail.

\subsection{Modeling \& Representation}
\label{sec:ModRep}

We encode a solution (\textit{chromosome}) as a set of move-class operations of variable lengths. More precisely, a solutions is a set of move-class operations $S =\{ C_i \rightarrow M_j, ..., C_h \rightarrow M_k\}$, where a generic
operation $C_i \rightarrow M_j$ denotes that the class $C_i$ is moved to the model $M_j$ (move-class operations). Note that, in this representation, a chromosome has variable size, i.e., the number of move-class operations can vary during the search (by adding or removing operations).

The choice to model the solutions as changes instead of a complete dependency structure (as done by~\citet{Harman2002}) was made to reduce the cost of evaluating the solution (fitness evaluation). Re-evaluating entire new proposals on how to modularize a software system takes significantly more processing power than solely evaluating a set of changes (i.e, the diffs). In our experience, representing solutions as a set of changes also dramatically reduces the memory consumption.
Besides, modeling solutions as set of changes has been previously proposed by Abdeen et al.~\cite{Abdeen2013} and also applied by other researchers in the field~\cite{Mahouachi2018, Mkaouer2016}.

\subsection{Search Objectives}
We optimize the following four search objectives:
\medskip

\textbf{Coupling}. This objective is often proposed as code quality measures~\cite{Meyers2007, Marcus2008, Jermaine1999, Hitz1996, Briand1998, Bavota2013, Bansiya2002}, and is often used in combination with search algorithms to improve the code quality of a code base (e.g., ~\cite{Abdeen2013, Ouni2012, Mahouachi2018, Praditwong2011, Mkaouer2015, Mitchell2006, Mahouachi2013, Harman2002, Chong2013, Bavota2012interact, Allen2001, Alizadeh2018}). In this paper, we measure software coupling using the \textit{inter-module coupling} (InterMD). InterMD measures the total number of dependencies modules have with the other modules of the system. A dependency between two modules $A$ and $B$ happen whenever any class from module $A$ depends on any class from module $B$. The overall InterMD of an individual is therefore measured as the sum of the module dependencies for all modules in the system.
\medskip

\textbf{Cohesion}. We use the \emph{intra-module coupling} (IntraMD) as the cohesion metric. IntraMD measures the number of intra-module dependencies (i.e., number of dependencies among classes within the same module) divided by the maximum possible amount of intra-dependencies, similar to
\citet{Harman2002}. We have experimented with two other metrics, based on the Common Closure Principle (CCP) and the Common Reuse Principle (CRP)~\cite{Martin2003}. Our preliminary analysis showed better improvements when using IntraMD, and therefore, it is the one we report in this paper.
\medskip

\textbf{Number of changes}. This objective measures the amount of ``move-class'' refactorings, i.e. the number of classes that have been moved to a different module in the solution. Therefore, given two solutions $X_1$ and $X_2$ achieving the same InterMD and IntraMD, we prefer the solution with fewer changes. This objective also reflects the \textit{manual effort} needed by developers to manually validate the generated solutions (refactoring recommendations) and eventually accept the changes. As we state before, many improvements can be made to a real-world large-scale code base; however, the effort needed might not be worth the gain in quality, nor companies are ready to risk large numbers of changes at once. 

\medskip

\textbf{Cost of module cache breaks}.
As mentioned before, the transitive coupling is an important and critical objective to consider for our industrial partner. \adyen uses module caching to avoid the recompilation of the entire code base. Caching is often used to remedy the fact that building code after every change can take a long time. 
Instead of re-compiling all the software system modules, \adyen only re-compiles modules that changed and their dependencies, recursively. 
This causes any change to have a cascading effect throughout the code base, in the reverse direction of the module dependencies. Removing a single dependency or splitting a module significantly effect on the number of caches broken. To the best of our knowledge, no prior work considered the cost of module cache breaks.

An adequate re-modularization should minimize the re-build time needed due to transitive coupling by reducing the number of module cache breaks. In theory, the simplest solution would be to directly measure the building time (with caching) for every candidate solution, i.e., at fitness evaluation time. Despite its theoretical simplicity, this approach is not feasible for large systems for two main reasons. First, properly measuring the building time would require to perform the build stage multiple times to have a reliable measure. Second, each building run requires almost one hour for our industrial system.

To overcome these limitations, we \textit{approximate the building cost} by considering on (1) the number of module cache breaks (which determines the number of modules to rebuild) and (2) the size of the module being modified.
More precisely, we \textit{estimate the cost of module cache breaks as follows}:
\begin{equation}
Cost(S) = \sum_{M_i \in M}  \mathcal{F}_{LOC}(M_i) * Cb(M_i)
\end{equation}
\noindent where $S$ is a candidate solution; $M$ is the set of modules in the system; $\mathcal{F}_{LOC}$ translates the size of the module (lines of code) into the corresponding build cost; $Cb(M_i)$ is the number of times the module cache is broken based on the rate of changes for the module $M_i$ itself and all its transitive dependencies.

$\mathcal{F}_{LOC}$ approximates the build cost of a module using a linear regression model that uses the lines of code (LOC) of a module and predicts its build time. 
We note that using $\mathcal{F}_{LOC}$ the estimated build time can be calculated quickly and can be easily re-evaluated after changes are made by simply subtracting the LOC of the moved classes from their original modules and adding it to the ones they were moved to. We show the linear-regression relation between time and LOC of a module and our regression function in Figure~\ref{fig:regression}.

\input fig-regression

\subsection{The search algorithm}
The Non-Dominated Sort Genetic Algorithm II (NSGA-II) \cite{Deb2002} was chosen to optimize this problem. This choice is supported by the fact that the number of search objectives to optimize is low ($<5$). Previous studies have shown that NSGA-II is the most effective algorithm for software re-modularization and similar problems  \cite{Abdeen2013, Alizadeh2018, Bavota2012interact, daSilva2020, Ghannem2017, Mkaouer2016, Mahouachi2018, Ouni2012}.

NSGA-II is a multi-objective evolutionary algorithm that iteratively optimizes an \textit{initial pool} of $N$ randomly generated solutions, i.e., sets of move-class operations in our cases. In each iteration, the fittest solutions are selected for using the \textit{binary tournament selections}. Selected pairs of solutions are genetically recombined using \textit{crossover} and \textit{mutation}. Newly generated solutions ---often referred to as to \textit{offspring}--- are evaluated to compute the objective scores. Finally, the $N$ \textit{parents} and $N$ \textit{offspring} solutions are inserted within the same pool, from which only $N$ solutions survive to the next generation. The $N$ solutions to survive are selected using the \textit{fast non-dominated sorting} and the \textit{crowding distance}~\cite{Deb2002}. The sorting aims to promote solutions that are more optimal than others based on the \textit{Pareto optimality}~\cite{Deb2002}. The crowding distance aims to promote well-diversified solutions within the phenotype space. The iterative process terminates when the maximum number of iterations is reached~\cite{Deb2002}.

In the following, we described the main elements of NSGA-II that we customized to our problem.
\medskip

\textbf{Population Initialization}. 
To generate the initial population, we create a pool of randomly generated solutions. Each random solution $S$ is created by creating $k$ move-class operations (delta changes). The number $k$ is randomly generated within the interval $[1; 100]$. In other words, our initial population will have solutions with different numbers of move-class operations.
\medskip

\textbf{Crossover}.
We propose a crossover operator that aims to preserve building blocks, which in this case are multiple changed classes that only have a positive effect when moved together. This operator is based on the crossover operator proposed by Harman et al.~\cite{Harman2002}. 

The \textit{uniform crossover} creates two offspring solutions by exchanging genes (move-class operations) between the two selected parents. The crossover does not exchange classes within the same module across the two parents to preserve building blocks. Instead, it copies all classes withing the same module all together into of the two offspring. This means that all classes in the same module of one parent are all copied in the same in one child, and the classes in the same module in the other parent in the other child. This is repeated for each module, pairing the two children with the two parents randomly for each.
\medskip

\textbf{Mutation}. 
We propose two mutation operators: (1) the \textit{min-cut-based operator} and (2) the \textit{neighborhood-based operator}.

The \textit{min-cut-based operator} picks a random class $C$ and builds its dependency graph. In such a graph, the nodes are classes, while the oriented edges denote that one class A (source node) depends on another class B (target node). The graph is built by first adding the randomly selected class $C$ and then all classes on which $C$ is directly control dependent. This process is repeated to add also the classes $C$ is transitively dependent on. The graph build process ends when a certain \textit{depth} is met, which is limited due to the computation cost associated with the min-cut. The mutation operator then performs the Stoer-Wagner algorithm to determine the min-cut of this graph \cite{Stoer1997}. The initially chosen class is in one of the resulting sub-graphs. All classes withing the sub-graph are then moved to a different module. This operator is inspired by existing research that used graph cutting algorithms for modularization \cite{Chong2013, Jermaine1999, Marx2010}.

The \textit{neighborhood-based operator} moves a pool of classes within the same module into another module. This operator was introduced by Fraser and Arcuri~\cite{Fraser2013} for test case generation. 
The mutation operator was adopted to our problem as follows: it picks a random class, adds it to a set $\mu$, then keeps adding random classes in the neighborhood until the condition $r>(c)^{n}$ is met. In the condition above, $r$ is a random number between 0 and 1; $c$ is a constant between 0 and 1 and impacts the average size of the resulting set; $n$ is the size of the set. The neighborhood of a class $C$ corresponds to all classes that are in the same module of $C$ and that have at least one structural dependency with the classes in the set $\mu$. This set is then moved to a different module.

\subsection{Circular dependencies} 
\label{constr}

Solutions that contain any circular dependency between two modules (e.g., module $A$ depends on module $B$, and module $B$ depends on module $A$) are invalid. 

During the design phase, we experimented with punishing and repairing solutions that violates the constraints (circular dependencies) similar to work of Deb et al.~\cite{Deb2002}. However, we have empirically observed that solutions with circular dependencies had naturally higher fitness values given their larger number of transitive dependencies. Therefore, we decided not to apply any punishment in case a solution contains a circular dependency. We nevertheless remove any possible solution with circular dependencies from the final set.

\subsection{Parameter Tuning}
\label{sec:tuning}
We identify the best hyper-parameters after running the algorithm in a reduced version of the code base that contained 17 modules and around 10k classes. We selected the 17 modules manually as to increase their generalizability towards the entire code base.

We make use of default values based on existing research and the intuition we built throughout its development. Given the large number of possible combinations and the computational cost required to run them all, we opt for tuning a smaller set of parameters, and to tune one parameter at a time while using a default value for the others. We run each configuration three times and picked the one with highest average value.

The best parameter values are as follows:

\begin{itemize}
\item Population size = 500.
\item Mutation probability $p_m=0.50$.
\item The two mutation operators (\textit{min-cut-based operand} and \textit{neighborhood-based}) have equal (and mutual exclusive) probability to be applied. In other words, each operator has 50\% chance to be applied.
\item Duplicates are deleted.
\item The crossover operator that preserves building blocks is used with a crossover probability $p_c=1.00$.
\item An elite archive with the same size as the population is used.
\end{itemize}

%% file: fig-regression.tex
\begin{figure}
\centering
\includegraphics[width=0.8\columnwidth]{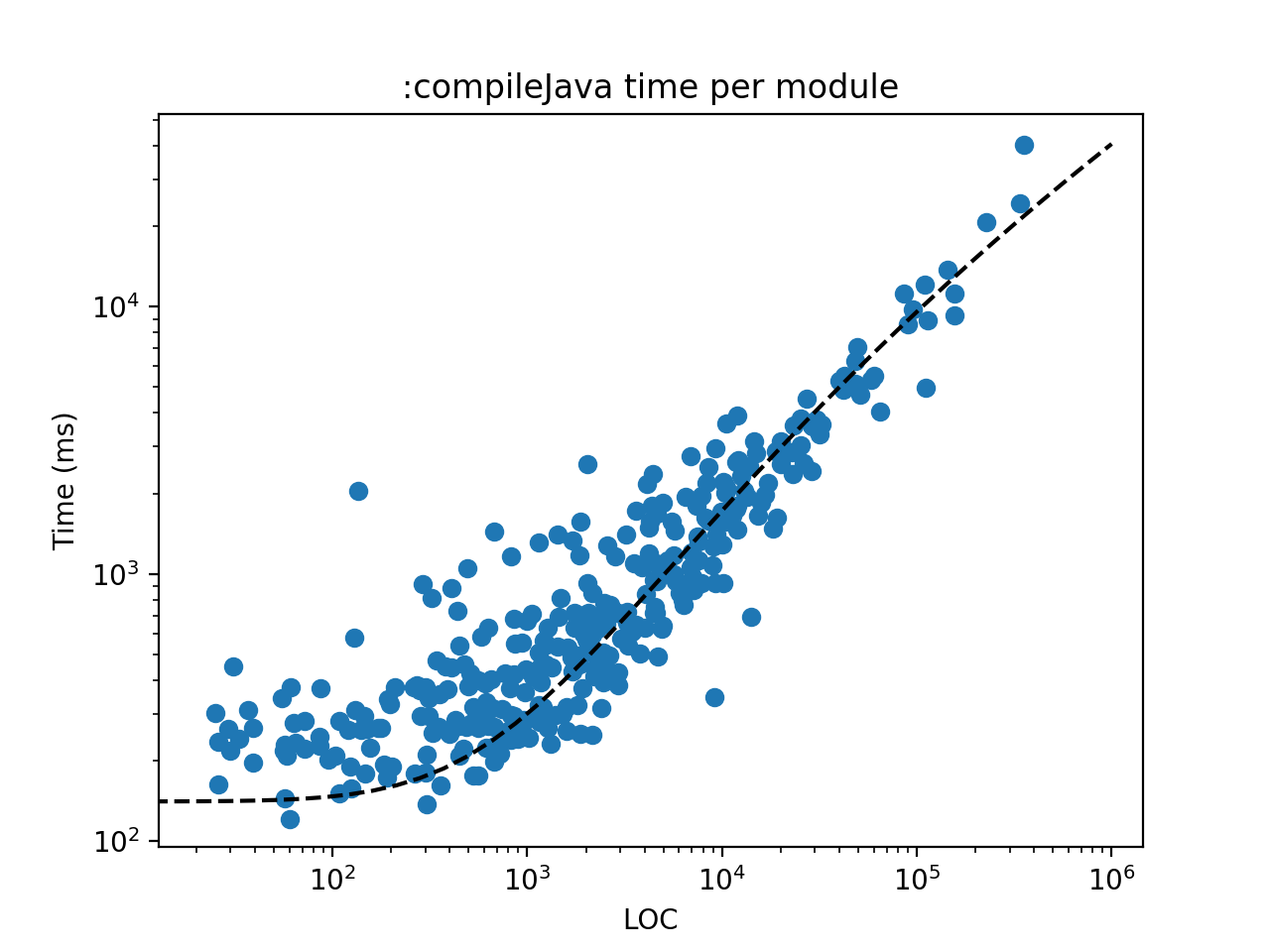}
\caption{The runtime of the compilation time and LOC per modules, and the regression function.}
\label{fig:regression}
\end{figure}

%% file: sec-research-part1.tex
\section{Study I: Effectiveness of the approach}

We set out our first research goal to measure the effectiveness of the proposed search-based algorithm in identifying solutions that would improve \adyen's code base. To that aim, we propose two research questions:

\newcommand{\rqA}{How effective is our approach in producing Pareto-efficient trade-offs that optimize the search objectives?}
\newcommand{\rqB}{How fast does our approach converge?}

\begin{description}
    \item[RQ$_1$:] \textit{\rqA}
    \item[RQ$_2$:] \textit{\rqB}
\end{description}

We run our approach on the entire \adyen code base for 24 hours on 8 cores at 2.1GHz with 8GB of RAM, two times, one for 24 hours, and one for 72 hours. 
We could not run NSGA-II more than once for each different search budget due to constraints imposed by the industrial context. Although multiple runs may generate slightly different results, we performed preliminary experiments for parameter tuning (see Section~\ref{sec:tuning}). We observed good stability of the results in our preliminary runs.

\subsubsection{Performance metrics}

Many performance metrics have been proposed in the literature to assess the optimality and distribution of the produced trade-offs~\cite{zitzler2003performance}. The most common performance metrics include hyper-volume (HV)~\cite{zitzler1999multiobjective}, inverted generational distance (IGD)~\cite{coello2004study}, spread~\cite{li2009spread}, and $R^2$~\cite{hansen1994evaluating}. However, these metrics requires knowing the true optimal Pareto front and measuring some quality aspects of the non-dominated front generated by NSGA-II (or any many-objective search algorithm). Some quality metrics compute the distance between the generated and true Pareto fronts (generational distance), diversity of the solutions in the front (spread), etc. However, in our context, the true Pareto front is unknown; thus, existing metrics like HV cannot be computed.

Another potential option could consist of creating a hybrid front (also called \textit{reference front}) by combining the non-dominated fronts obtained by using multiple MOEAs, and over multiple runs. This was also not a viable solution as we could not run multiple MOEAs nor performing more repetitions due to the industrial context constraints. 

Based on the observation above, we answer RQ$_1$ by analyzing the maximum improvements that can be obtained in the search objectives. This allows us to measure the best improvements in the software quality metrics, i.e., the extreme points of the produced Pareto front. This also corresponds to the \textit{regions of interests} (ROI) of our industrial partner, whose goal was to understand the extreme alternatives (trade-offs) across the selected search objectives.
To answer RQ$_2$, we analyze how the best improvements in the software quality metrics change over time. More precisely, we analyze the extreme points of the Pareto front every 50 generations, based on the 72 hours run.

\subsection{RQ1: \rqA}

\input fig-rq1
\input tab-rq1

The results of the runs done to answer RQ1 can be seen in Figure \ref{fig:rq-1}, which provides a graphical representation of the Pareto front of both runs. Table \ref{tab:rq1-results} reports the relative improvements of the quality metrics (search objectives).

\observation{Solutions identify significant improvements, as measured by the optimization metrics.} In the most fit solutions (i.e., the Pareto front), we observe improvements of 205.1\% in terms of cohesion (IntraMD), 3.6\% in terms of coupling (InterMD), and 1.6\% in terms of transitive dependencies and cache breaks (EBCCB). If we focus in single search objectives (i.e., disregarding the quality in the other objectives), we observe even larger potential improvements for the specific metrics, such as 530.4\% for IntraMD, 6\% for InterMD, and 1.6\% for EBCCB.

\observation{Different optimal solutions need different amounts of changes.} We see that the extreme solutions in the front require from 67 up to 170 class moves to be implemented. Interestingly, the solution that maximizes EBCCB seems to need the least amount of changes, while the solution that maximizes IntraMD needed the most. For the IntraMD metric, we observe that solutions that maximize it tend to suggest creating many new small modules (i.e., modules with a few classes only), which explains why optimizing IntraMD requires many class moves. On the other hand, single class moves may already drastically improve the EBCCB metric, e.g., a class move that removes a large number of transitive dependencies in the graph, explaining why solutions with fewer changes are enough to have a high impact in the metric.

\subsection{RQ2: How fast does the proposed algorithm converge?}

\input fig-rq2

Figure~\ref{fig:rq-2} shows the improvement of quality metric (search objectives) over the generations during the 72-hours run.

\observation{The algorithm seems to converge for the InterMD and the EBCCB metrics.} The algorithm convergence at generation 19000 for the InterMD metrics; for the EBCCB, the algorithm seems to be mostly converged, only finding minor improvements after generation 14000. Indeed, we can observe that both metrics reach a plateau in Figure~\ref{fig:rq-2}.

\observation{72 hours was not enough to observe the algorithm converging for the IntraMD metric.} 
We conjecture that the algorithm has not converged yet because the optimal solution for this metric is a structure constructed of modules with one class per module. The mutation operators can easily move a single class to a new module, which keeps improving this metric over generations. This process will continue until the one class per module structure is achieved. However, this solution would overfit our problem as developers would not accept it. Part of our future work includes avoiding this behavior by, for example, adding more objectives that conflict with IntraMD.

%% file: fig-rq1.tex

\begin{figure}
\centering
\includegraphics[width=0.9\columnwidth]{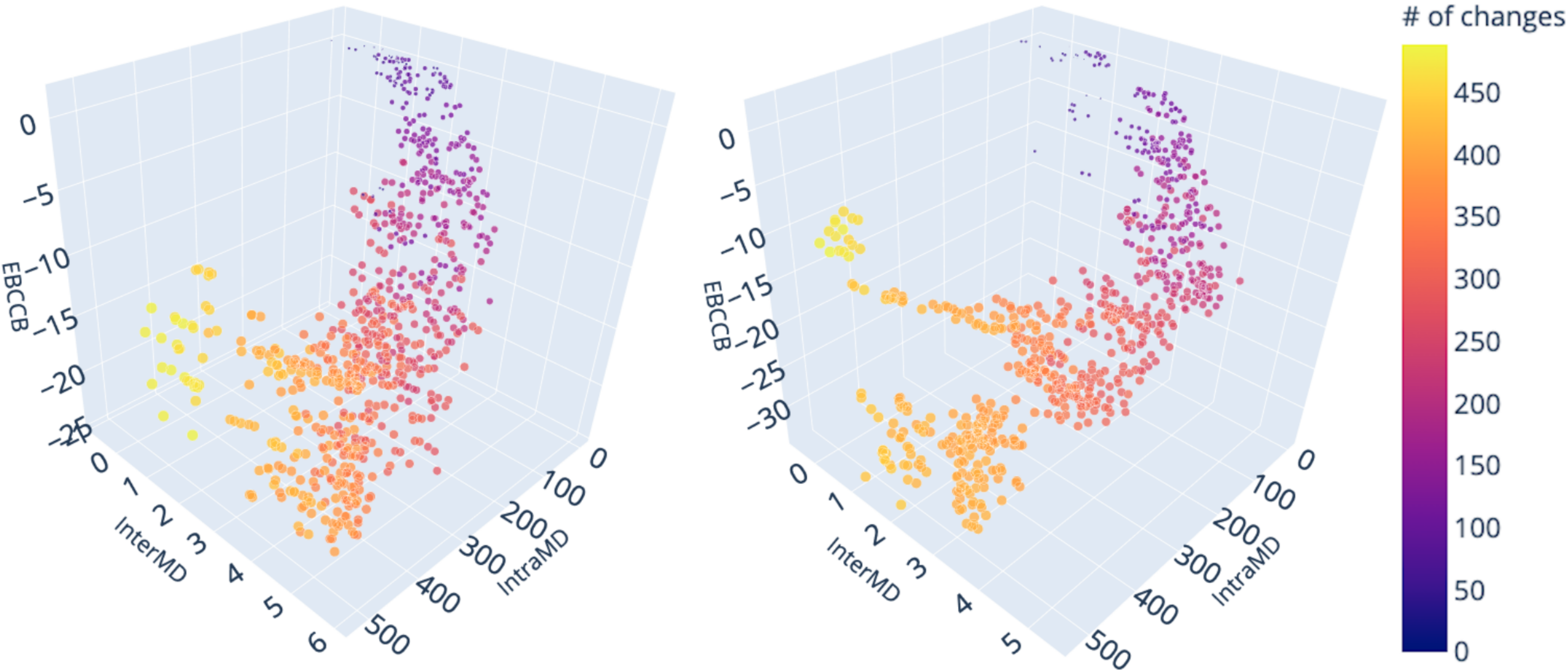}
\caption{Representation of the Pareto front of the first (left) and the second (right) runs.}
\label{fig:rq-1}
\end{figure}

%% file: tab-rq1.tex


\begin{table}
\centering
\caption{Solutions in the Pareto front. The IntraMD metric does not grow linearly.}
\label{tab:rq1-results}
\begin{tabular}{lrrr}
\toprule
 & \textbf{Best IntraMD} & \textbf{Best InterMD} & \textbf{Best EBCCB} \\ \midrule
\multicolumn{3}{l}{\textbf{Optimizing all variables}} \\
\midrule
IntraMD & 205.1\% & 151.3\% & 55.0\% \\
InterMD & 3.5\% & 3.6\% & 1.2\% \\
EBCCB & 0.1\% & 0.9\% & 1.6\% \\
\# of moves & 170 & 138 & 67 \\ \midrule
\multicolumn{3}{l}{\textbf{Optimizing single variables}} \\ \midrule
IntraMD & 530.4\% & 322.3\% & 55.0\% \\
InterMD & 1.2\% & 6.0\% & 1.2\% \\
EBCCB & N/A & -8.1\% & 1.6\% \\
\# of moves & 418 & 244 & 67 \\ \bottomrule
\end{tabular}
\end{table}

%% file: fig-rq2.tex
\begin{figure}
\centering
\includegraphics[width=0.8\columnwidth]{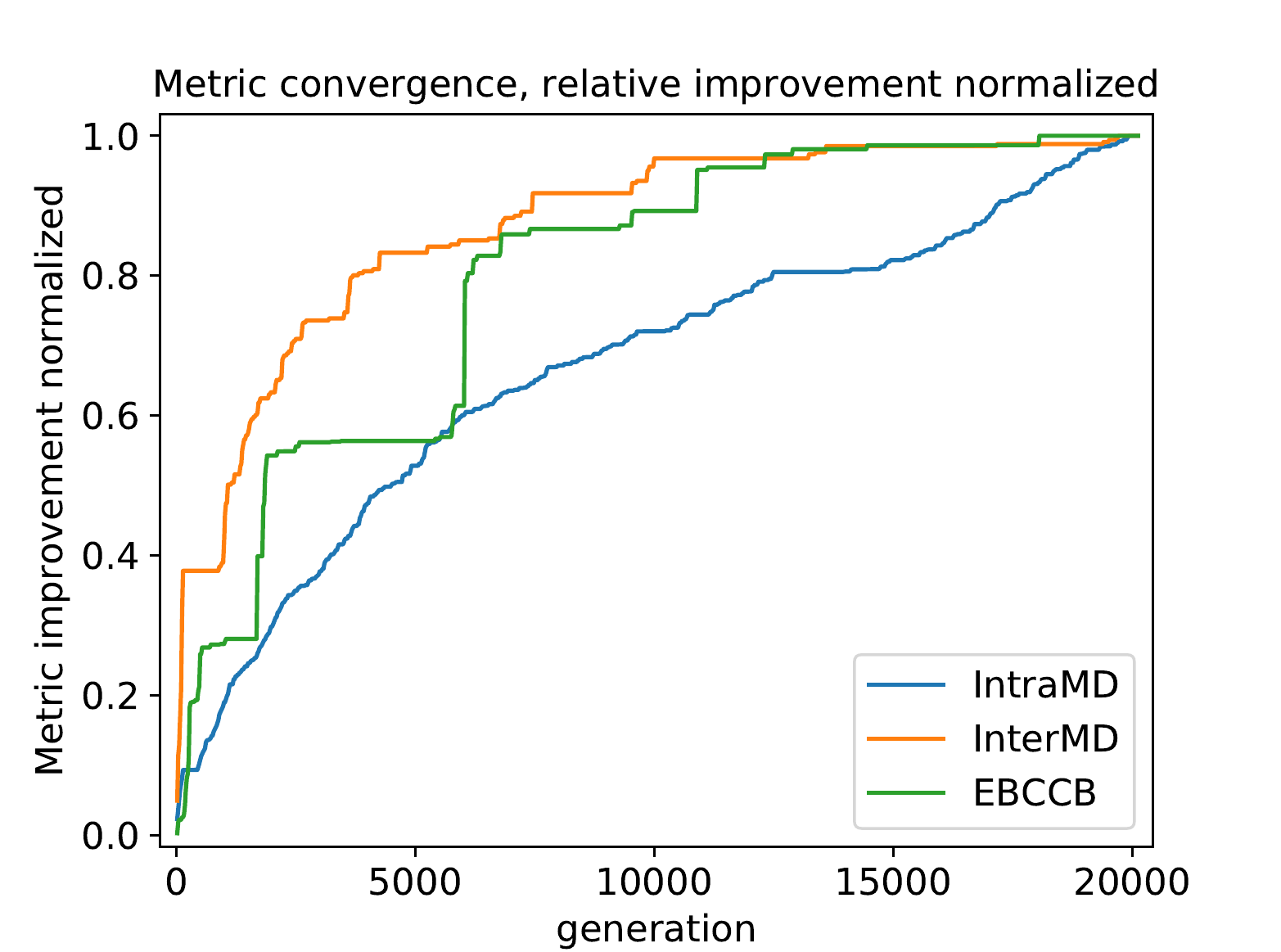}
\caption{The fittest solution of the IntraMD, InterMD, and EBCCB over the 72-hours run. Number of changes is not represented in the graph as it is always near zero.}
\label{fig:rq-2}
\end{figure}

%% file: sec-research-part2.tex
\section{Study II: Developers' perceptions}

As further contribution of our paper, we aim to understand whether the solutions proposed by the approach are considered useful by the developers at \adyen. To that aim, we propose another research question:

\newcommand{\rqC}{How do developers perceive the re-modularization suggestions of the approach?}

\begin{description}
    \item[RQ$_3$:] \textit{\rqC}
\end{description}

\subsection{Methodology}

We take groups of refactoring suggestions from the (trade-off) solutions produced by NSGA-II in the previous step of this research; then, we present these solutions to the developers at \adyen responsible for that module and capture their perception on how useful that particular recommendation is.
Given that solutions in the Pareto front improve the entire code base, and our developers work on single modules (and thus, no single developer can review the entire solution), we broke down solutions according to the modules they improve on. For example, suppose any solution X suggests improvements in module A and module B; we consider the refactoring operations related to module A and the refactoring operations related to module B as two possible suggestions to be reviewed by developers. Moreover, we select suggestions that had between 1 to 10 class moves (i.e., a solution small enough that developers could review during their work time) and that improved on at least 2 of the 3 optimization metrics (i.e., IntraMD, InterMD, and EBCCB) and did not worsen the other. We make an exception for the InterMD metric, which was allowed to be worsened if the EBCCB was improved, as InterMD counteracts the process of creating new modules and the EBCCB improvement hints at an improvement in the overall coupling. This process resulted in 13 recommendations to be reviewed by the developers.

We choose a developer who has made multiple changes to the refactoring recommendation classes over the past 6 months to review the suggestions. We leverage \adyen's code repository to extract such information. If no such developer exists or can be interviewed, we choose a senior developer related to that module.
This resulted in 11 developers judging 13 different suggestions. Table \ref{devtable} reports the experience of the 11 participants on \adyen's code base and in general code development. Four of the developers were seniors instead of developers that made changes to the classes.

During the interview, we asked each participant to review the re-modularization suggestion s/he was assigned to, based on their experience and contributions (see Table \ref{devtable}). We further asked the participant to assess the merit of the re-modularization suggestion.
If the suggestion was considered good, we asked the interviewee to rate the importance of the change. If the suggestion was considered not good, we then ask the developer to judge whether that class is indeed in the right place and whether it should be moved to another module.

\input tab-rq2-developers

\subsection{Results}

In Table~\ref{tab:rq2-result}, we show a summary of the developers' perceptions, whether the identified problematic class is indeed considered problematic, and whether the recommended module for the class to be moved to is considered accurate. 

\input tab-rq2-results

\observation{All the 13 suggestions revealed possible improvements to the code base.} The developers considered all classes in the 13 suggestions to be currently placed in the wrong module and that they should be moved elsewhere. One exception happens in change C5, where one developer did not consider the recommendation as accurate.

\observation{Around half of the suggestions of where to move the classes were deemed to be accurate or partially accurate by developers.} Developers labeled seven out of the 13 suggestions about where to move the classes as accurate or partially accurate. Four of these recommendations could be implemented directly. The remaining suggestions were not fully accurate. Among the reasons gave by the developers, we highlight:
(i) classes were at the right modules but were poorly designed in terms of coupling or cohesion;
(ii) the classes are dead code or superseded by other functionality, so they should be deleted instead;
(iii) future plans concerning the classes hinder the suggested change;
and (iv) the module the classes are moved to are already in the process of being split up.

\observation{Developers did not approve any new proposed modules.} Interestingly, developers did not approve any suggestions for creating a new module. Here, we argue that this might be caused by the size of the changes the developers analyzed. Given that the changes were small in size, the suggested new modules were too granular (i.e., contained only a few classes). Therefore, we do not fully disregard the idea of the algorithm proposing new modules, and we will study the developers' perceptions on new modules in future work.

\observation{Different developers make different judgments and conclusions.} We observe that judging the quality of a recommendation is indeed prone to the developers' perceptions. For example, in the case of C5, two developers gave completely contradictory judgments on the suggestion. D4 determined C5 to be bad and saw no possible improvements, while D7 judged it to be a good suggestion. Moreover, D2 and D6 had similar reasons when judging C3 and C7, but end up with different conclusions. In both cases, the suggestion would be good in the current state of the code, but given the company's future plans related to the feature that the module implements, D2 concluded that the refactoring should be implemented and reverted when needed, whereas D6 concluded that it should not be implemented. Overall, this shows that deciding whether or not a suggestion is good depends on several factors. While optimizing numbers helps developers understand the overall benefit that the change will bring to the code base, in future work, we intend to study the decision-making process to better support developers in such challenging decisions.


%% file: tab-rq2-developers.tex
\begin{table}
\centering
\caption{The interviewed developers and their experience. Developers that were chosen as seniors to review a suggestion are marked with (s). Experience is represented in years.}
\label{devtable}
\begin{tabular}{lrrr}
\toprule
\textbf{Developer} & \textbf{Experience as} & \textbf{Experience at} & \textbf{Reviewed} \\
 & \textbf{a developer} & \textbf{Adyen} & \textbf{Suggestions} \\
\midrule
D1 & 9 & 4 & C1 (s), C2 (s) \\
D2 & 3.5 & 3.5 & C3 \\
D3 & 10 & 2.5 & C4 \\
D4 & 10 & 0.5 & C5 \\
D5 & 20 & 7 & C6 (s) \\
D6 & 6 & 3 & C7, C8 \\
D7 & 3 & 3 & C5 \\
D8 & 10 & 2 & C9, C10 (s) \\
D9 & 34 & 1.5 & C11 (s) \\
D10 & 2.5 & 2.5 & C12 \\
D11 & 6 & 2.5 & C13 \\ \bottomrule
\end{tabular}
\end{table}

%% file: tab-rq2-results.tex
\begin{table}[]
\centering
\caption{The developers' perceptions. The first column indicates whether developers considered the identified classes to be problematic. The second column indicates whether the suggestion of modules that classes should be moved to is accurate. Developers fully agree (\checkmark), partially agree (\texttildelow), or fully disagree (x) with the suggestion.}
\label{tab:rq2-result}
\begin{tabular}{lrr}
\toprule
\textbf{Change ID}     & \textbf{Considered a flaw?} & \textbf{Accurate move suggestion?}  \\
    \midrule
C1  & \checkmark                  & x                               \\
C2  & \checkmark                  & \texttildelow					\\
C3  & \checkmark                  & \checkmark                      \\
C4  & \checkmark                  & \texttildelow                   \\
C5  & \checkmark / x              & \checkmark / x 	                \\
C6  & \checkmark                  & x                               \\
C7  & \checkmark                  & x                               \\
C8  & \checkmark                  & \checkmark                      \\
C9  & \checkmark                  & x                               \\
C10 & \checkmark                  & x                               \\
C11 & \checkmark                  & \checkmark                      \\
C12 & \checkmark                  & x                               \\
C13 & \checkmark                  & \texttildelow                   \\ \bottomrule
\end{tabular}%

\end{table}

%% file: sec-challenges-ooportunities.tex
\section{Research Challenges and Opportunities}

Our results show that search-based re-modularization approaches scale to enterprise-level code bases, and solutions show significant relative improvements.
Nevertheless, we still see room for improvements and potential research directions. In the following, we highlight the challenges and opportunities we observed throughout this research, hoping that they will pave the road for future research.

\challenge{The need for better cohesion metrics.}
Despite IntraMD being the most used search objective to improve the cohesion of software modules, in practice, we observed that the algorithm can ``overfit'' by moving a single class to a new module multiple times, increasing the metric value significantly without actually improving the codebase. This also means that the metric also will not converge until a solution of one class per module has been constructed. For these reasons, we now believe that IntraMD does not accurately represent a given module or code base's cohesion. We, therefore, suggest future work to explore better cohesion metrics. We conjecture that the developers did not accept many move-class recommendations due to this type of overfitting.

In our preliminary study, we have explored the Common Closure Principle (i.e., classes that change together are packaged together) and the Common Reuse Principle (i.e., classes that are used together are packaged together), as proposed by \citet{Martin2003}, and so far experimented solely by \citet{Abdeen2013}. We observed fewer improvements during the parameter tuning phase, and thus, we discarded them in that stage. Future studies should investigate whether the solutions we obtain using these metrics while providing less relative improvements, do in fact, improve the cohesion of modules.

Finally, given the advances of natural language processing techniques and the similar statistical properties that source code hold~\cite{hindle2012naturalness}, we suggest future work to explore whether semantically similar classes should also be packed together. We observe researchers taking the first steps towards such an idea (i.e., \citet{Ouni2012}, \citet{Mkaouer2015}, \citet{Mahouachi2018}); we reinforce the need for such approaches.

\challenge{The need for more precise measures of effort.}
The effort (and risk) of introducing large changes in large-scale software systems should always be taken into account in any provided recommendation. In this research,
we used the number of move-class operations as a proxy for effort. This may not properly represent the effort required for the changes, as some classes might be significantly harder to move than others. For example, moving a class to a different module may imply re-designing the test code base, transferring the code ownership to a new team, etc. 

We argue that more realistic metrics to measure the cost of applying the refactoring in practice is needed. We identified a single attempt to improve the effort measurement (i.e., the MoJoFM metric~\cite{Wen2004}, which has been used by \citet{Bavota2012interact}). We suggest future work to explore, together with developers, what metrics best reflect the real effort of performing any re-modularization refactoring.

\challenge{The need for more domain knowledge in the process.}
We have proposed the EBCCB metric to capture the impact of transitive dependencies in \adyen's module cache mechanism. The metric seems to converge. Not many changes are required to achieve significant improvements in the code base. Code change suggestions that mostly improved on the metric were validated by developers. 

These results show the importance of adding domain knowledge, which may be company-specific, to the algorithm. We suggest future research to empirically look for other domain-specific functions that can be aggregated to the process. Given that most research in the area is conducted in open-source systems, without real access to developers, we did not find any paper proposing specific search objectives, which we consider as a clear opportunity for future work.

\challenge{The need for humans-in-the-loop.} We also have observed that relative improvements that the solutions bring to the code base may not be perceived as positive by the developers. We believe that any feedback given by the developers should be fed back to the algorithm. While we have not tried it in this research, we noticed that capturing the developers' perceptions for any proposed recommendation is an expensive task for the company. In this study, developers had to stop working on their main tasks and focus on our research. We envision a more streamlined way of capturing this information in the future. More specifically, we plan to serve recommendations during code review time. Whenever a developer touches a class that is considered to be in the wrong place by our approach, the tool would alert that developer about it. The tool would then ask the developer whether the recommendation is accurate and how important it is to fix it. The tool would show the same recommendation to several developers before not showing it anymore. Their perceptions would then be used back in the next search, i.e., ignore classes that developers consider to be in the correct place. 

Integrating these recommendations in their development life cycle would help developers fix the recommendation sooner and provide us with relevant data for more longitudinal empirical studies. We are only aware of the work from \citet{Bavota2012interact}, which made use of interactive GA. We suggest future approaches to consider such interaction.

%% file: sec-ttv.tex
\section{Threats to Validity}


\textit{Internal validity.}
We did not explore every combination of parameters during parameter tuning. Exploring every combination or testing every possible parameter value would result in a vast amount required runs, making the optimization infeasible. In other words, the used combination may not be optimal. More extensive parameter tuning may result in better performance and potentially better solutions from the algorithm.

The suggested groups of class move refactorings that were shown to developers in the interview process are not entirely representative of the solutions they originate from, due to the filtering and grouping process beforehand. Due to this, the conclusion that all suggested groups indicate flaws in the code might not hold for all the class move refactorings that are contained in the entire solutions. Nevertheless, we argue that the portions we show to developers are a significant part of the solution, and thus, serve as a good representative for the entire solutions.



\textit{External validity.}
In regards to external validity, the approach in this paper was applied to a (large) code base which is used and developed by a single company. More research is required to determine whether the results are generalizable to systems using different programming languages, developed by different people, and fulfilling different purposes.

%% file: sec-conclusions.tex
\section{Conclusions}
\label{cha:conc}

The re-modularization of large-scale software systems is a challenging and expensive task. Developers often have little information about the (positive or negative) impact that changes in modules may have in the software architecture of their systems.

In this paper, we perform a case study on the effectiveness of search-based approaches for software re-modularization at \adyen, a large-scale enterprise software system, composed of million of lines of code and hundreds of developers. 
We also propose a new optimization variable that takes transitive coupling and building time as part of the search problem; a variable that is perceived as fundamental in practice by the \adyen developers. 

Our results show that such approaches can scale to large code bases and are able to identify modules of the system that require improvements. Nevertheless, recommendations as to which module to move the class to still requires improvement. We hope this paper highlights the importance of such line of research and provides researchers with more insights on how these techniques currently behave in large-scale industrial software systems.